\documentclass[aps,pra,preprint,superscriptaddress,showpacs]{revtex4-1}
\usepackage[T1]{fontenc}
\usepackage[latin2]{inputenc}
\usepackage{color}
\usepackage{amsmath}
\usepackage{amssymb}
\usepackage{graphicx}
\usepackage{setspace}
\usepackage{subfig}
\PassOptionsToPackage{version=3}{mhchem}
\usepackage{mhchem}

\makeatletter

\@ifundefined{showcaptionsetup}{}{%
 \PassOptionsToPackage{caption=false}{subfig}}
\usepackage{subfig}
\makeatother

\begin{document}

\title{Towards the full quantum dynamical description of photon
induced processes in \ce{D2+}}

\author{A. T\'oth}
\affiliation{Department of Theoretical Physics, University of Debrecen,H-4010 Debrecen, PO Box 5, Hungary}

\author{S. Borb\'ely}
\email[]{sandor.borbely@phys.ubbcluj.ro}
\affiliation{Faculty of Physics, Babe\c{s}-Bolyai University,Kog\u{a}lniceanu Street 1,400084 Cluj Napoca, Romania}

\author{G. Zs. Kiss}
\affiliation{Faculty of Physics, Babe\c{s}-Bolyai University,Kog\u{a}lniceanu Street 1,400084 Cluj Napoca, Romania}

\author{G. J. Hal\'asz}
\affiliation{Department of Information Technology, University of Debrecen, H-4010 Debrecen, PO Box 12, Hungary}

\author{\'A. Vib\'ok }
\email[]{vibok@phys.unideb.hu}
\affiliation{Department of Theoretical Physics, University of Debrecen,H-4010 Debrecen, PO Box 5, Hungary}
\affiliation{ELI-ALPS, ELI-HU Non-Profit Ltd,H-6720 Szeged, Dugonics t\'{e}r 13, Hungary}

\date{\today}

\begin{abstract}
A new quantum dynamical model has been developed to describe
the dissociative
ionization of deuterium molecular ions by intense laser pulses ($\tau=10$ fs, $\lambda=200$
nm and $I=3\times10^{13}$ W/cm$^{2}$). We calculated the ionization
probability  densities  by solving the time-dependent Schr\"{o}dinger equation numerically. 
Throughout the simulation the nuclear vibration was
considered as a dynamic variable with fixed molecular axis orientation.
Benchmark calculations were performed for the ionization of \ce{HeH++},
for which accurate numerical results are available in the literature,
in order to check the performance of this new restricted model.

\end{abstract}


\maketitle

\section{Introduction}

The dynamics initiated in a molecule by photon impact is usually
discussed in terms of the Born-Oppenheimer (BO) approximation
\cite{Born-Oppenheimer}, where fast moving electrons are treated
separately from the slow nuclei. In this scheme, electrons and nuclei
do not easily exchange energy. However, this energy exchange can 
become important in some nuclear regions, particularly
in the vicinity of degeneracy points or conical intersections (CIs) \cite{Horst1,Baer1,Graham1,Wolfgang1,Baer2,Matsika1}.
It is widely recognized today that conical intersections are very
important in the nonadiabatic processes which are ubiquitous in photophysics
and photochemistry. 

For diatomic molecules, which have only one degree of freedom, it
is not possible for two electronic states of the same symmetry to become degenerate and, 
as a consequence of the well-known noncrossing rule, an avoided crossing results.
However, this is only true in free space. It has been shown in previous papers that conical intersections
can be created even in diatomics \cite{Nimrod1,Nimrod2} both by running or by standing
laser waves. In this situation the laser-light couples either the center of mass motion with
the internal rovibrational degrees of freedom (in case of a standing
laser field) or the vibrational motion with the emerged rotational
degree of freedom (in case of a running laser field) resulting in a so-called
light-induced conical intersection (LICI). The position of
the LICI is determined by the laser frequency, while the laser intensity controls
the strength of its nonadiabatic coupling. 

A few years ago, we have started a systematic study of the nonadiabatic
effect induced by laser fields in molecular systems and demonstrated
that the light-induced conical intersections have a significant impact
on several dynamical properties (like the molecular spectra,
the molecular alignment etc.) of diatomic molecules \cite{Gabor1,Gabor2,Gabor3,Gabor4,Gabor5,Gabor6,Gabor7,Gabor8}.

The photodissociation and ionization processes of the \ce{D2+}
ion have been thoroughly studied for more than quarter of a century
\cite{Schumacher1,Bandrauk1,Bandrauk2,Bandrauk3,Charron1,Bandrauk4,Charron2,Atabek1,Sanding,Atabek2,Posthumus,Atabek3,Uhlmann1,Wang1,Anis1,Anis2,Esry1,Agnes1,Paul1,Thum1,Fischer1,Fischer2,Anis3,He1,Furakawa,He2,Litvinyuk,Thumm1,Esry2,Fischer,Lefebvre,Alejandro1,Corkum,Fojon,Guan,Yuan,Handt,Pavicic1,Pavicic2,Selsto,Fernandez,Kelkensberg,Picon,Bainbridge,Nabekawa,Khosravi,XuHan,Wanie}.
In our recent papers, we have also investigated the dissociation dynamics
of the \ce{D2+} in the LICI picture starting the initial
nuclear wave packet either from different vibrational eigenstates
or from the Franck--Condon distribution of the vibrational states
obtained from photoionizing \ce{D2} \cite{Gabor5,Gabor6,Gabor7,Gabor8}.
One (1D) and two-dimensional (2D) calculations have been performed
and compared to each other. In the 1D model, the molecular rotational
angle was only a parameter, i. e. the LICI was not considered, whilst
in the 2D scheme the rotational angle was assumed as a dynamic variable 
thereby explicitly incorporating the LICI. 
The obtained 1D and 2D results strongly differ, demonstrating
the significant impact of the LICI on the \ce{D2+} dissociation dynamics. 
Additionally, in a recent letter \cite{Gabor8},
we were able to provide the first ``direct observable and measurable
signature'' of the light-induced conical intersections 
by studying carefully the dissociation process of \ce{D2+}.

This article is our initial attempt to combine the photodissociation and ionization of \ce{D2+} 
using a newly developed restricted model which assumes fixed molecular axis orientations.
Beside the previously considered \cite{Gabor5,Gabor6,Gabor7,Gabor8} ground and first excited states,
this model includes an ionized state with a well defined asymptotic electron momentum $\vec{k}$.
For this three-level system, the time-dependent nuclear Schr\"{o}dinger equation was solved 
numerically and the ionization probability density for the fixed $\vec{k}$ electron momentum 
was extracted from the final population of the ionized state.
The ionization probability density for the whole electron continuum was mapped by 
performing several independent calculations for the different electron momentum values.

Using the above outlined approach, we have investigated the ionization spectrum of \ce{D2+} 
for previously studied conditions \cite{Gabor5,Gabor6,Gabor7,Gabor8}, where the formation of 
the LICI between the ground and the first excited state has been observed. Our model has 
also been used to investigate the ionization of \ce{HeH++} at fixed molecular axis lengths, 
for which accurate numerical results are available in the literature \cite{Guan,Fojon}.

This article is divided in four sections. The introduction is followed by 
the presentation of the required methods and algorithms as well as the calculated dynamical quantities. 
The third section presents and discusses the \ce{D2+} and the \ce{HeH++}  numerical results.
In the last section the conclusions and future plans are summarized.

\section{Theory and methods }

\begin{figure}
\begin{centering}
\includegraphics[width=0.45\textwidth]{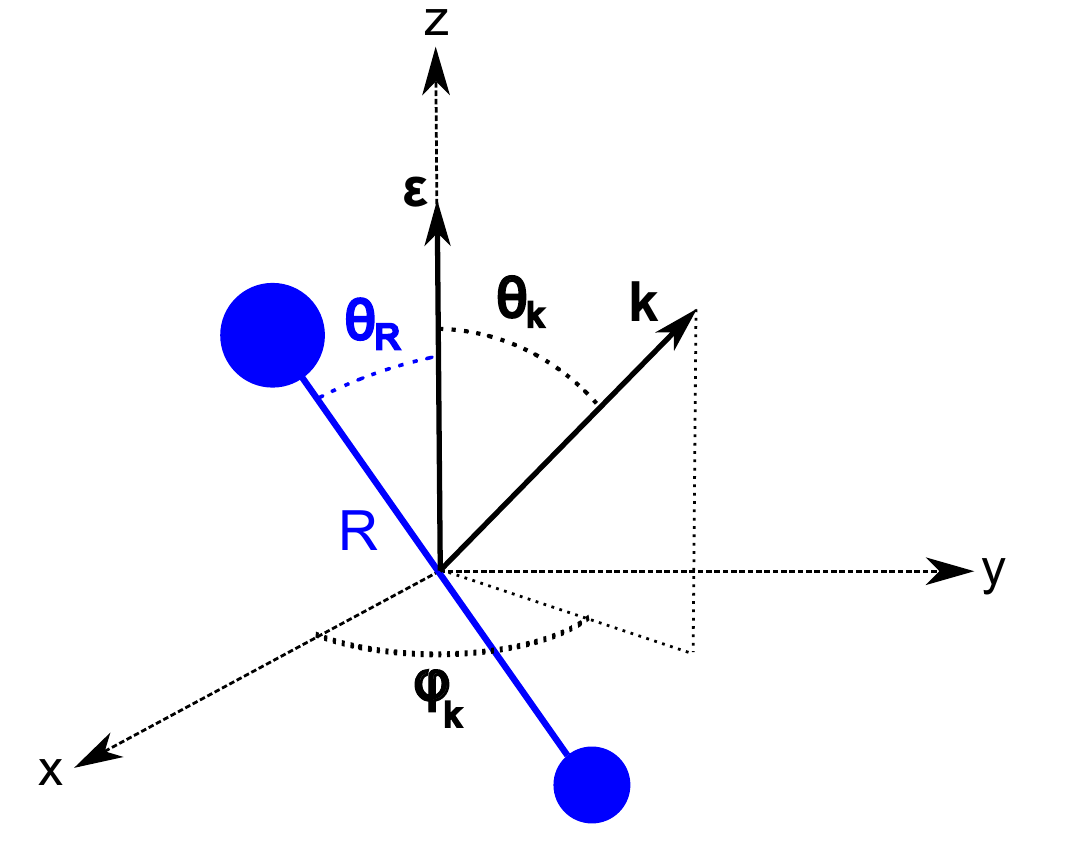} 
\end{centering}

\caption{The geometry of the studied system in a reference frame defined by
the laser polarization ($\varepsilon\in Oz$) and by the molecular
axis ($R\in xOz$). The ejected electron is described by the $\vec{k}$
asymptotic momentum vector.}

\label{fig:geometry} 
\end{figure}

Our theoretical approach for laser driven ionization and dissociation dynamics 
is an extension of a previous model that described the dissociation
of \ce{D2+} molecular ion \cite{Gabor5,Gabor6,Gabor7,Gabor8}.
The original model describes nuclear dynamics only for
the two relevant electronic energy levels corresponding to the ground
and the first excited states. The extended model includes a
third energy level corresponding to an electronic continuum
state defined by its asymptotic electron momentum $\vec{k}$.
The wave packets contributing to the nuclear wave function for this
model system can be arranged in a vector form as follows 
\begin{equation}
\Psi_{N}(R;\theta_{R})=\left(\begin{array}{c}
\Psi_{G}(R;\theta_{R})\\
\Psi_{X}(R;\theta_{R})\\
\Psi_{I}(R;\theta_{R})
\end{array}\right),\label{eq:ansatz}
\end{equation}
where $\Psi_{G}(R;\theta_{R})$ is the vibrational wave packet on
the ground state, $\Psi_{X}(R;\theta_{R})$ is the wave packet on
the excited state, while $\Psi_{I}(R;\theta_{R})$ is the wave packet
on the ionization level. $R$ denotes the internuclear separation
between the nuclei, while $\theta_{R}$ is the angle between the internuclear
axis and the polarization axis of the external laser pulse. The geometrical arrangement
of the studied system is shown in Figure \ref{fig:geometry}. In the current
calculations, $\theta_{R}$ is considered only as
a parameter and the nuclear wave packet dynamics is solved explicitly
only for the vibrational degree of freedom. 
The field free Hamiltonian of the system can be written as
\begin{equation}
H_{0}=\left(\begin{array}{ccc}
-\frac{1}{2\mu}\frac{\partial^{2}}{\partial R^{2}}+V_{G}(R) & 0 & 0\\
0 & -\frac{1}{2\mu}\frac{\partial^{2}}{\partial R^{2}}+V_{X}(R) & 0\\
0 & 0 & -\frac{1}{2\mu}\frac{\partial^{2}}{\partial R^{2}}+V_{I}(R)
\end{array}\right),
\end{equation}
where $\mu=M_{1}M_{2}/(M_{1}+M_{2})$ is the reduced mass of the system
and $V_{\{G,X,I\}}(R)$ are the potential energy surfaces of the ground
($G$), the first excited ($X$) electronic, as well as the ionic
($I$) states of the \ce{D2+} ion. These potential energy surfaces
can be expressed as a sum of the electronic energy and Coulomb repulsion
between the atomic nuclei: 
\begin{equation}
V_{\{G,X,I\}}(R)=E_{\{G,X,I\}}(R)+\frac{Z_{1}Z_{2}}{R},
\end{equation}
with $Z_{1}$ and $Z_{2}$ being the charge of the nuclei. 
\begin{figure}
\begin{centering}
\includegraphics[width=0.45\textwidth]{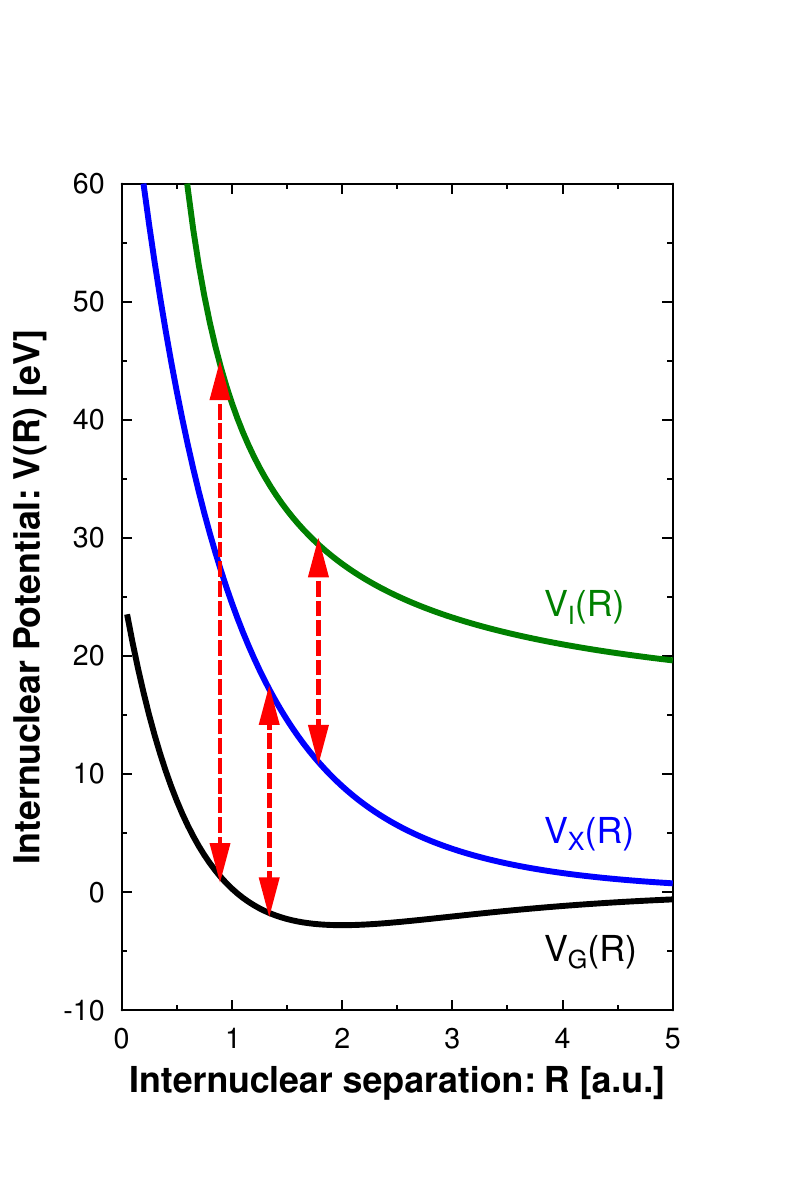} 
\par\end{centering}

\caption{A schematic picture of the potential energy surfaces considered in
the present model showing three energy levels: the ground $V_{G}(R)=1s\sigma_{g}$,
the first excited $V_{X}(R)=2\pi\sigma_{u}$ and the ionization level
$V_{I}(R)$ (with $k=0.2$ a.u.) of \ce{D2+}.
The radiative couplings between the different energy levels
are indicated by dashed arrows. }

\label{fig:ensurf} 
\end{figure}

Figure \ref{fig:ensurf} shows that the origin of the potential
energy axis is fixed to the asymptotic value of the ground state potential
energy surface. Coupling between the different levels is possible
in the presence of an external laser pulse and is described
by the interaction term of the Hamiltonian: 
\begin{equation}
H_{int}=\left(\begin{array}{ccc}
0 & -\vec{\varepsilon}\cdot\vec{d}_{GX}F(t) & -\vec{\varepsilon}\cdot\vec{d}_{GI}F(t)\\
-\vec{\varepsilon}\cdot\vec{d}_{XG}F(t) & 0 & -\vec{\varepsilon}\cdot\vec{d}_{XI}F(t)\\
-\vec{\varepsilon}\cdot\vec{d}_{IG}F(t) & -\vec{\varepsilon}\cdot\vec{d}_{IX}F(t) & 0
\end{array}\right),
\end{equation}
where $\vec{d}_{ij}(R,\theta_{R})=\left\langle \psi_{i}^{e}(\vec{R},\vec{r})|\vec{r}|\psi_{j}^{e}(\vec{R},\vec{r})\right\rangle $
are the transition dipole moments between the electronic states described
by the $\psi_{i}^{e}$ and $\psi_{j}^{e}$ wave functions and $F(t)$
is the electric component of the employed laser pulse. 
Sine-square shaped laser pulses are used throughout the calculations:
\begin{equation}
F(t)=\left\{ \begin{array}{ll}
F_{0}\cos(\omega t)\sin^{2}(\frac{\pi t}{\tau}) & \mathrm{if}\quad t\in(0,\tau)\\
0 & \mathrm{elsewhere}
\end{array}\right.,\label{eq:pulse}
\end{equation}
where $\omega$ is the carrier wave frequency, $F_{0}$ is the
amplitude of the electric field and $\tau$ is the duration of the
laser pulse. The time evolution of the nuclear wave function (\ref{eq:ansatz})
is governed by the time-dependent Schr\"{o}dinger equation:
\begin{equation}
i\frac{\partial\Psi_{N}(R,t;\theta_{R})}{\partial t}=(H_{0}+H_{int})\Psi_{N}(R,t;\theta_{R}).\label{eq:ntdse}
\end{equation}
In the present approach, this Schr\"{o}dinger equation (\ref{eq:ntdse}) is solved 
using the efficient multi configurational time-dependent
Hartree (MCTDH) package \cite{mctdh_1,mctdh_2,mctdh_3,mctdh_4,mctdh_4_1,mctdh_5}.
To describe the vibrational degree of freedom the 
Fast Fourier Transformation-Discrete Variable Representation (FFT-DVR) was used,
with $N_{R}$ basis elements distributed on the range from 0.1 a.u. to
$R_{\mathrm{max}}$. The vibrational wave function can be expressed as 
a function of this FFT-DVR basis set ($\chi_{j}(R)$) as:

\begin{equation}
\Psi(R,t)=\sum_{j=1}^{N_{R}}c_{j}(t)\chi_{j}(R).\label{eq:MCTDH-wf}
\end{equation}

A complex absorbing potential (CAP) is applied to all three energy levels considered in our model. 
It has the following form

\begin{equation}
-iW(R)=\begin{cases}
\begin{array}{cc}
0 & \mathrm{if}\\
-i\eta(R-R_{0})^{3} & \mathrm{if}
\end{array} & \begin{array}{c}
R<R_{0}\\
R\in[R_{0},R_{max}]
\end{array}\end{cases},
\end{equation}
where $\eta$ is the strength, and $R_{0}$ is the starting point
of the CAP. In our model, the CAP
prevents the reflection of the wave function's dissociative part at
the simulation box boundary and helps to monitor
the absorbed wave function norms at each energy level, which are
subsequently used to calculate physical observables. After initialization,
the nuclear wave packet (\ref{eq:ansatz}) is propagated numerically
in time according to the Schr\"{o}dinger equation (\ref{eq:ntdse})
using the 6th order Adams-Bashforth-Moulton (ABM) predictor-corrector
method with variable time-steps. The value of the numerical parameters
used during the numerical solution of the Schr\"{o}dinger
equation (\ref{eq:ntdse}) are presented along the results obtained
for each investigated system.

\subsection{The electronic Schr\"{o}dinger equation\label{sub:The-electronic-Schr=0000F6dinger}}

Beside the accurate solution of the nuclear time-dependent Schr\"{o}dinger
equation (\ref{eq:ntdse}) the most important part of the proposed
work is the accurate calculation of the electronic energy levels $E_{\{G,X,I\}}(R)$
and of the transition dipole moments between them. In order to achieve
this goal, the one-electron stationary Schr\"{o}dinger equation
for diatomic molecules is solved with the electronic Hamiltonian written as:
\begin{equation}
H^{e}=\frac{\hat{p}_{e}^{2}}{2}-\frac{Z_{1}}{r_{1}}-\frac{Z_{2}}{r_{2}},\label{eq:eham}
\end{equation}
where $r_{1}$ and $r_{2}$ are the electron coordinates measured
from the nuclei. From practical purposes it is more convenient to convert the electronic Hamiltonian (\ref{eq:eham}) into
prolate spheroidal coordinates: 
\begin{align}
 H^{e}=-\frac{2}{R^{2}(\xi^{2}-\eta^{2})}&\left[\frac{\partial}{\partial\xi}(\xi^{2}-1)+\frac{\partial}{\partial\xi}\frac{\partial}{\partial\eta}(1-\eta^{2})\frac{\partial}{\partial\eta}+\frac{1}{\xi^{2}-1}\frac{\partial^{2}}{\partial\phi^{2}}+\frac{1}{1-\eta^{2}}\frac{\partial^{2}}{\partial\phi^{2}}\right]-\nonumber\\
 &\nonumber\\
 &-\frac{2}{R}\left[\frac{\xi(Z_{1}+Z_{2})+\eta(Z_{1}-Z_{2})}{\xi^{2}-\eta^{2}}\right],\label{eq:prham}
\end{align}
where $R$ is the fixed internuclear distance, while the $\eta$ and $\xi$ coordinates are obtained using the following transformations:
\begin{align}
\xi&=\frac{r_{1}+r_{2}}{R},\\
\eta&=\frac{r_{1}-r_{2}}{R}.
\end{align}

In the next step, the above Hamiltonian (\ref{eq:prham}) was discretized using a
finite-element discrete variable representation (FEDVR) numerical
grid \cite{FEDVR} for the $\eta$ and $\xi$ coordinates. The $\phi$
coordinate can be discretized on the $e^{im\phi}$ trigonometric basis.
In the present case we were looking for cylindrically symmetric
solutions, meaning that this basis could be reduced to $m=0$, therefore in (\ref{eq:prham})
the $\phi$ dependent terms were omitted altogether. The eigenvalues and eigenvectors
of the obtained Hamiltonian matrix were calculated using the SLEPc
\cite{slepc} eigensolver package. Further technical details on the above
outlined procedure can be found in \cite{zsolt:2014}.

After performing detailed convergence checks, the accurate electronic
wave functions ($\Psi_{\{G,X\}}^{e}(R)$) and electronic energies
($E_{\{G,X\}}^{e}(R)$) for the ground and excited states were obtained
at all internuclear distances considered during the nuclear
wave packet dynamics. The electronic energy for the ionization level is given by:
\begin{equation}
E_{I}(R)=\frac{k^{2}}{2}.
\end{equation}
 For the description of the ionization states, simple
one-center momentum normalized Coulomb wave functions \cite{messiah} have been used:
\begin{equation}
\psi_{c}(\vec{k},\vec{r})=\sqrt{\frac{2}{\pi}}\frac{1}{kr}\sum_{lm}i^{l}e^{i\sigma_{l}}Y_{lm}^{*}(\hat{k})Y_{lm}(\hat{r})F_{l}(\gamma,kr),
\label{eq:coulomb}
\end{equation}
where $F_{l}$ are the radial Coulomb functions, $\gamma=-(Z_{1}+Z_{2})/k$
is the Sommerfeld parameter, and $\sigma_{l}=\arg(\Gamma(l+1+i\gamma))$
is the Coulomb phase shift. As the Coulomb wave functions are ''state
density'' functions, they need to be discretized by dividing the momentum
space into small boxes with a $\delta V$ volume centered around the
$\vec{k}$ momentum vector. As a result of this discretization all
continuum states in this $\delta V$ volume element can be collectively
described by the $\psi_{I}^{e}(\vec{k},\vec{r})$ wave function, which
can be written as 
\begin{equation}
\psi_{I}^{e}(\vec{k},\vec{r})=\sqrt{\delta V}\psi_{c}(\vec{k},\vec{r}).
\end{equation}
The calculation of the potential energy surfaces and of the transition dipole moments 
is straightforward from these wave functions and their corresponding eigenenergies. 

\subsection{Calculation of observables}

After the end of the laser pulse (\ref{eq:pulse}), the system
is propagated further in time until on each energy level, the dissociative
part of the nuclear wave packets reached the end of the simulation
box, where they are absorbed by the CAPs. The norm of these absorbed
wave packets gives the probability of dissociation from each level.
The ionization level is a special case as the norm of the nuclear
wave packet at the end of the laser pulse also gives the 
probability of ionization into the $\delta V$ momentum volume element:
\begin{equation}
P(\delta V,\vec{k})=\left|\langle\Psi_{I}(R,t>\tau)|\Psi_{I}(R,t>\tau)\rangle\right|^{2}.\label{eq:ionprob}
\end{equation}
For all molecular systems studied here, the ionization level is purely
dissociative, thus the searched $P(\delta V,\vec{k})$ probability
is equal with the wave function norm absorbed by the CAP placed on
the ionization level. By dividing the probability (\ref{eq:ionprob})
with the $\delta V$ volume element the ionization probability density is obtained: 
\begin{equation}
P(\vec{k})=\lim\limits _{\delta V\rightarrow0}\frac{P(\delta V,\vec{k})}{\delta V}.\label{eq:probdens}
\end{equation}
In order to map this ionization probability density 
independent calculations for all relevant $\vec{k}$ values must be performed.

\subsection{Limitations of the model}

The main limitation of the above outlined model is rooted in its design:
the ionization probability densities for each ionization channel are
calculated separately. This implies that the direct and indirect (through
the ground and excited states) population transfer between the ionization
channels is not permitted. This shortcoming of our model may have
an important impact on the calculated ionization probability densities
at high laser field intensities where the continuum-continuum transitions
and the depletion of the ground state play an important role in the
ionization dynamics. As the model is designed to be used in
situations where the dissociation and ionization have a comparable
importance, we believe that this limitation will not
have a significant impact on the obtained results.

\section{Results and discussion}

The previously presented model has been used to study the ionization of \ce{D2+} molecular ion by ultrashort laser pulses. 
The laser pulse parameters were those from our previous calculations \cite{Gabor5,Gabor6,Gabor7,Gabor8}, where 
LICI formation between the ground and the first excited state has been observed and
multiphoton ionization is the dominant ionization mechanism.

The ionization of the \ce{HeH++} molecule at fixed molecular axis lengths in the single-photon ionization regime
has been investigated to benchmark the new model. This system has been chosen as there are high 
precision differential results available \cite{Guan,Fojon}, and the \textit{ab initio} 
results can be directly compared to simple first order perturbation calculations.

\subsection{The \ce{D2+} ion}

In this subsection, we show detailed analysis of the photoelectron 
distributions obtained by further ionizing a \ce{D2+} molecular ion. The initial 
nuclear wave packet was obtained by transferring the vibrational ground state 
of the neutral molecule to the ground electronic state of the ion. The dynamics 
of this Franck--Condon wave packet was driven by a 10 fs long laser pulse with a 
200 nm wavelength of the carrier wave and $3\times10^{13}$ W/cm$^{2}$ intensity.
We only considered fixed molecular axis orientations as this is the first step 
in working toward our long-term goal of the complete dynamical 
description of diatomic molecules. 

\subsubsection{Numerical details }

A simulation box with $R_{max}$=30 a.u. size and $N_{R}$=768 gridpoints was used
for the present calculations.
A CAP with $R_{0}$=20 a.u. and $\eta=0.00005$ was applied to the ground state, 
while a CAP with $R_{0}$= 25 a.u. and $\eta=0.00236$ was used on the excited and 
ionization states. These parameters allow an accurate representation of the
nuclear wave packets on the considered electronic states, and prevent
the absorption of the nuclear wave packets during the action of the
laser pulse. 

The electronic states and the transition dipole moments between them
were calculated by solving the electronic Schr\"{o}dinger equation directly,
as outlined in subsection \ref{sub:The-electronic-Schr=0000F6dinger}.
The numerical convergence of the calculated $\Psi_{G}(R)$ and $\Psi_{X}(R)$
was carefully verified and the derived potential energy surfaces
and the dipole transition moments were compared to data found in the
literature \cite{dipole,pot} and a good agreement was found.

\begin{figure}
\includegraphics[width=0.45\textwidth]{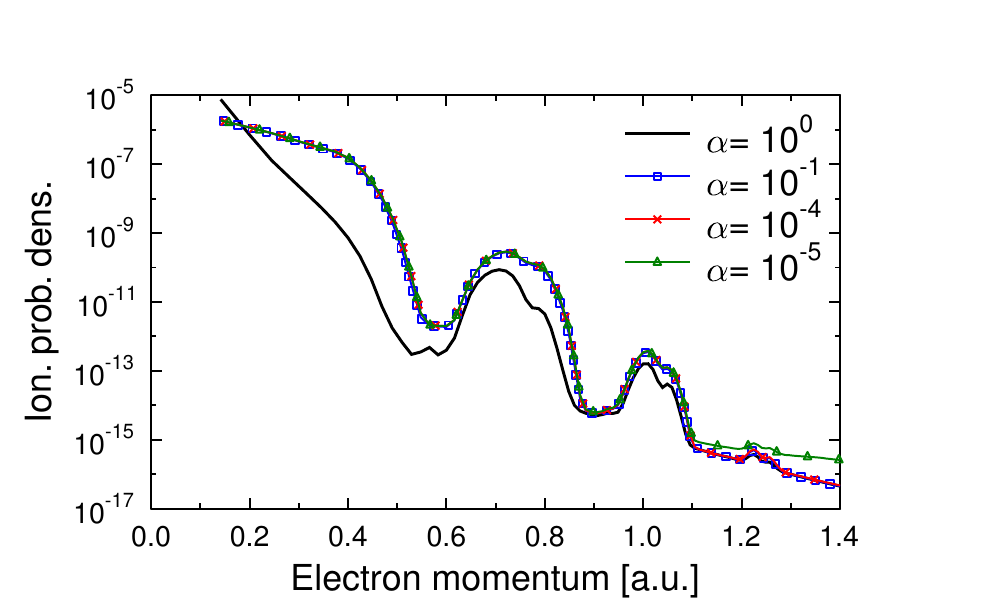}

\caption{Ionization probability density of the $\ce{D2+}$ along the internuclear
axis for different values of the $\alpha$ parameter. The molecular
axis is parallel to the laser polarization $\hat{\varepsilon}$. The
parameters of the driving laser field: $\lambda$ = 200 nm, $I_{0}$=3$\times$10$^{13}$
W/cm$^{2}$, $\tau$ = 10 fs. \label{fig:dvconv}}
\end{figure}

The correct selection of the $\delta V$ volume element used during the discretization of
the momentum space, and during the calculation of the ionization probability
densities (\ref{eq:probdens}) is an important component of the present approach.
In this model, we have fixed this volume element to be
\begin{equation}
\delta V=\alpha k^{2},\label{eq:volumeel}
\end{equation}
where $k$ is the magnitude of the electron momentum. The optimal value of the
$\alpha$ parameter was obtained from the convergence
tests performed for the laser pulse parameters used in the
calculations. The results of this convergence test are shown on Figure
(\ref{fig:dvconv}), where the ionization probability density for
electrons ejected along the molecular axis are shown for different
values of the $\alpha$ parameter. 

During the convergence tests, it was observed that a large $\alpha$ value 
leads to erroneous results for the low energy electrons
(see the $\alpha=1$ line) while a diminutive $\alpha$ value (see the $\alpha=10^{-5}$
line) leads erroneous results for high energy electrons. Despite this,
a large interval for $\alpha$ {[}$\in(10^{-1},10^{-4})${]} was found over 
which the obtained ionization probability converges and
$\alpha=0.001$ was chosen to be used throughout all calculations.

\subsubsection{Electron spectra}

Calculations for different molecular axis orientations
($\theta_{R}\in\{0^{\circ},30^{\circ},60^{\circ},90^{\circ}\}$) have been performed 
for a fixed laser pulse with the following parameters: $\lambda$= 200 nm, 
$I_{0}$=3$\times$10$^{13}$ W/cm$^{2}$, $\tau$ = 10 fs.
\begin{figure}
\includegraphics[width=0.45\textwidth]{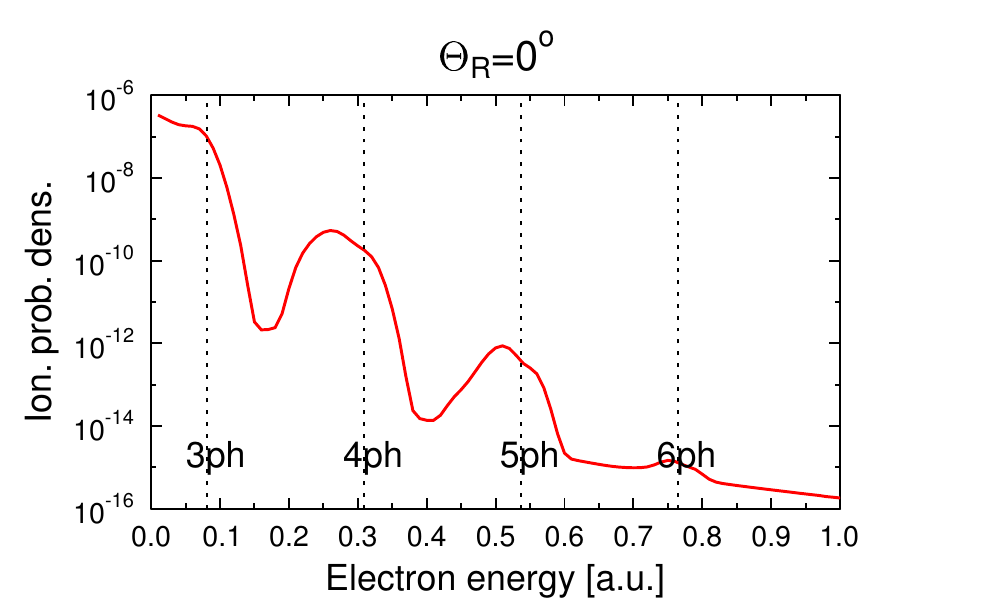}

\caption{Angle integrated ionization probability density
as a function of electron ejection energy for the $\theta_{R}=0^{\circ}$
molecular axis orientation. The laser pulse parameters are $I_{0}$=3$\times$10$^{13}$
W/cm$^{2}$, $\tau$ = 10 fs. With vertical dashed lines the 3-6 photon
nonsequential transition thresholds measured from the bottom of the $V_{G}(R)$
potential energy surface are indicated.
\label{fig:endep_00}}
\end{figure}
Figure (\ref{fig:endep_00}) shows the 
angle integrated ionization probability density as a function of electron
ejection energy for $\theta_{R}=0^{\circ}$ molecular axis orientation.
The 3-6 photon transition thresholds measured from the bottom
of the $V_{G}(R)$ potential energy surface have been plotted with dashed vertical lines.
These indicate the location of the non-sequential multiphoton ionization peaks.

\begin{figure*}
 \begin{minipage}{0.45\textwidth}
 \subfloat[First peak (E=0.08 a.u.)]{\includegraphics[width=\textwidth]{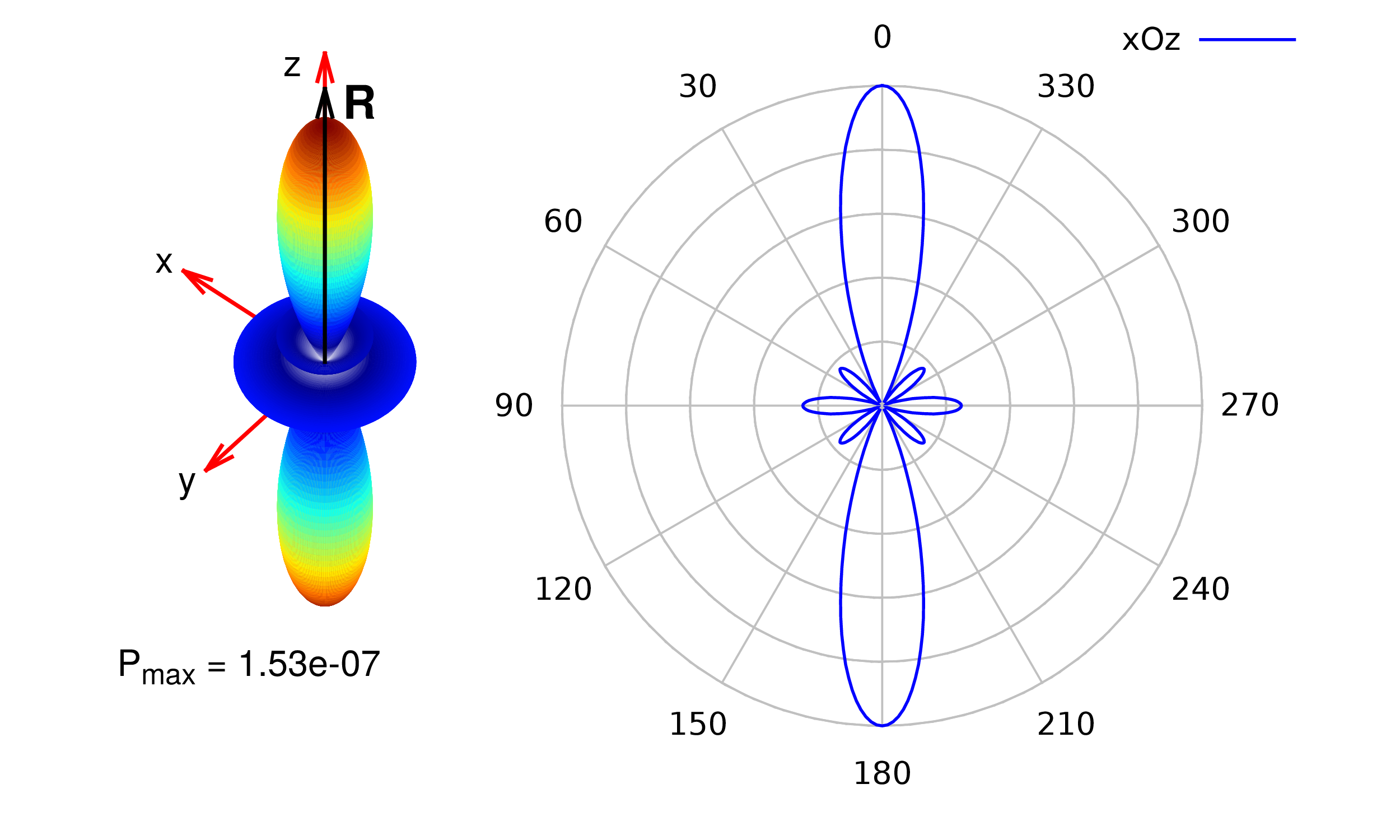}}
 \end{minipage}
 \begin{minipage}{0.45\textwidth}
 \subfloat[Second peak (E=0.31 a.u.)]{\includegraphics[width=\textwidth]{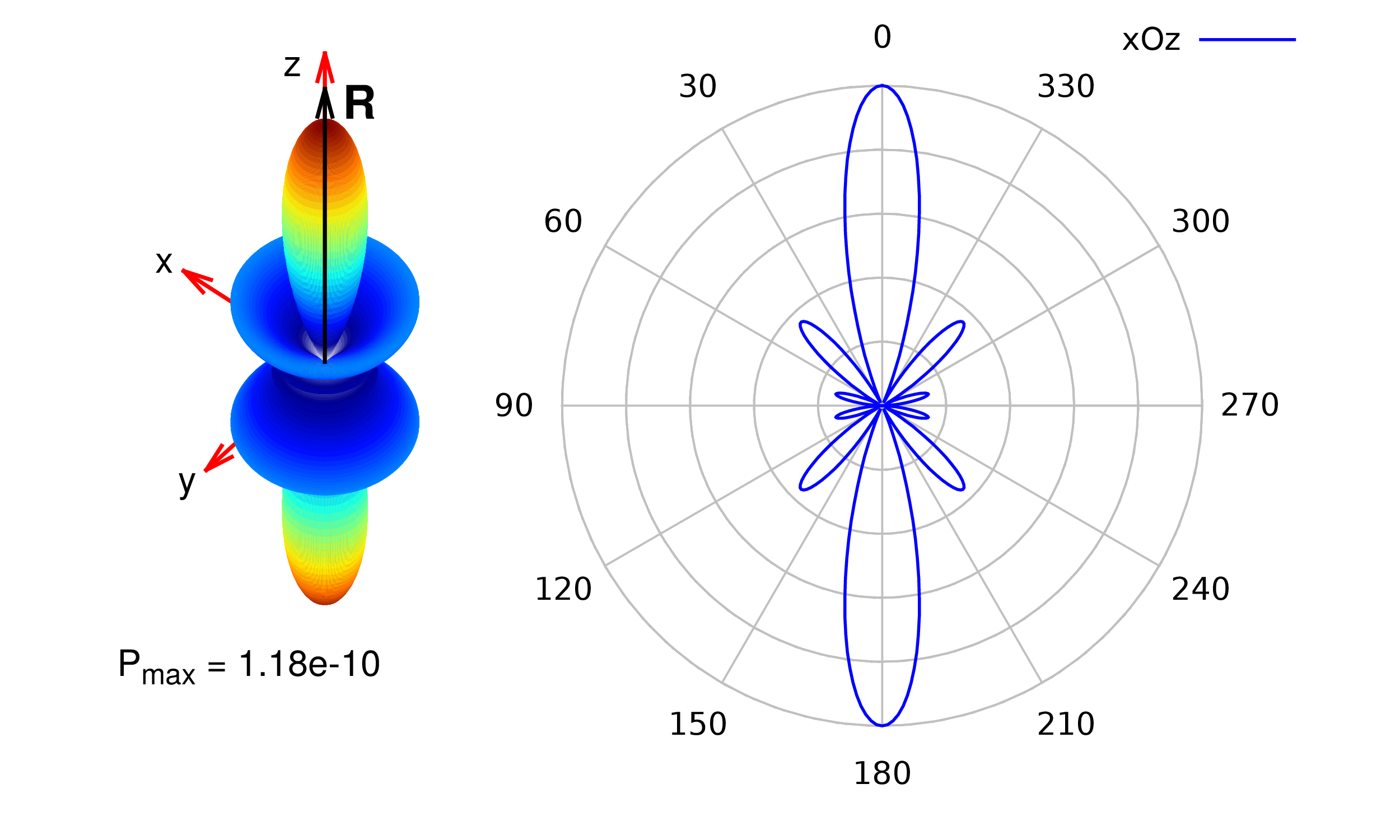}}
 \end{minipage}
  \begin{minipage}{0.45\textwidth}
 \subfloat[Third peak (E=0.54 a.u.)]{\includegraphics[width=\textwidth]{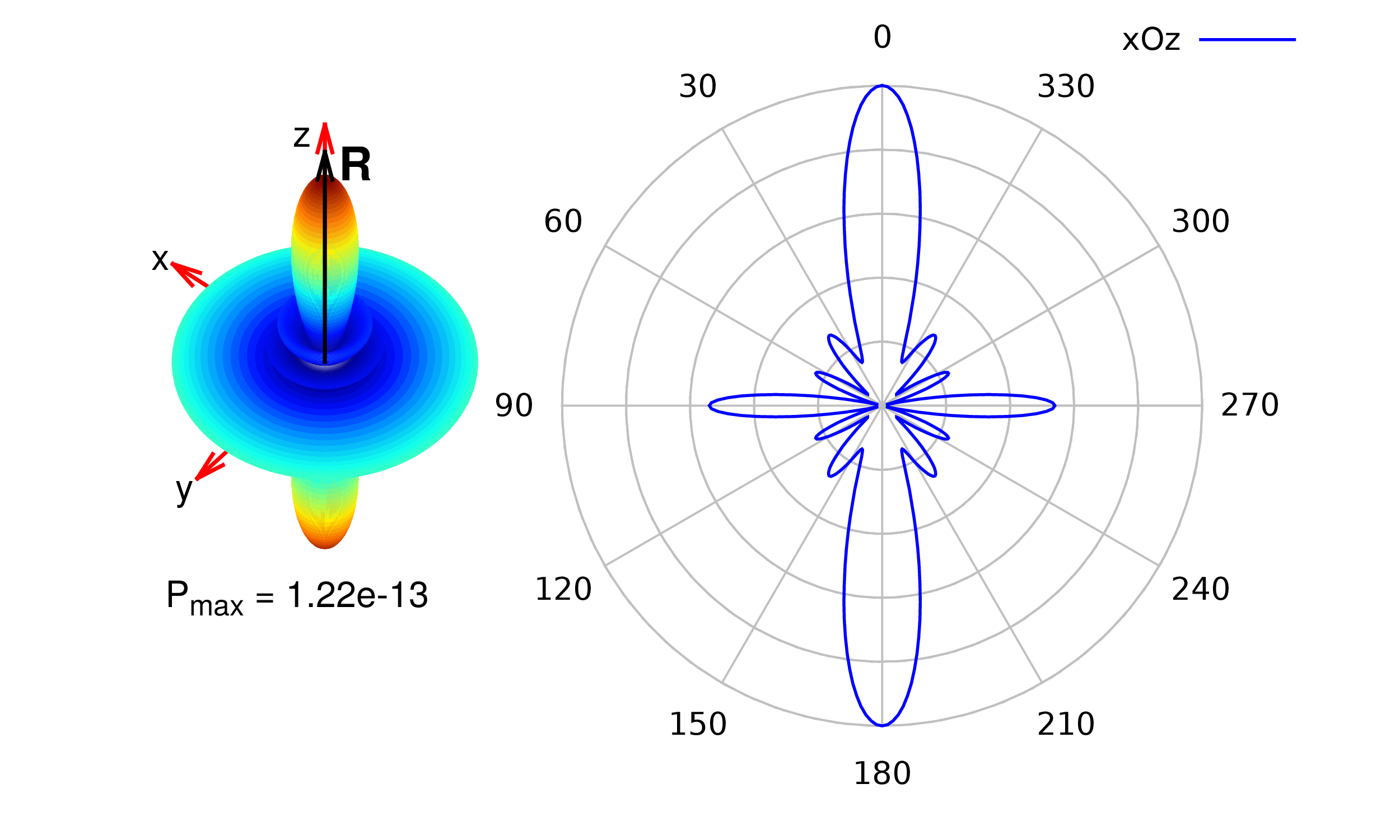}}
 \end{minipage}
  \begin{minipage}{0.45\textwidth}
 \subfloat[Fourth peak (E=0.76 a.u.)]{\includegraphics[width=\textwidth]{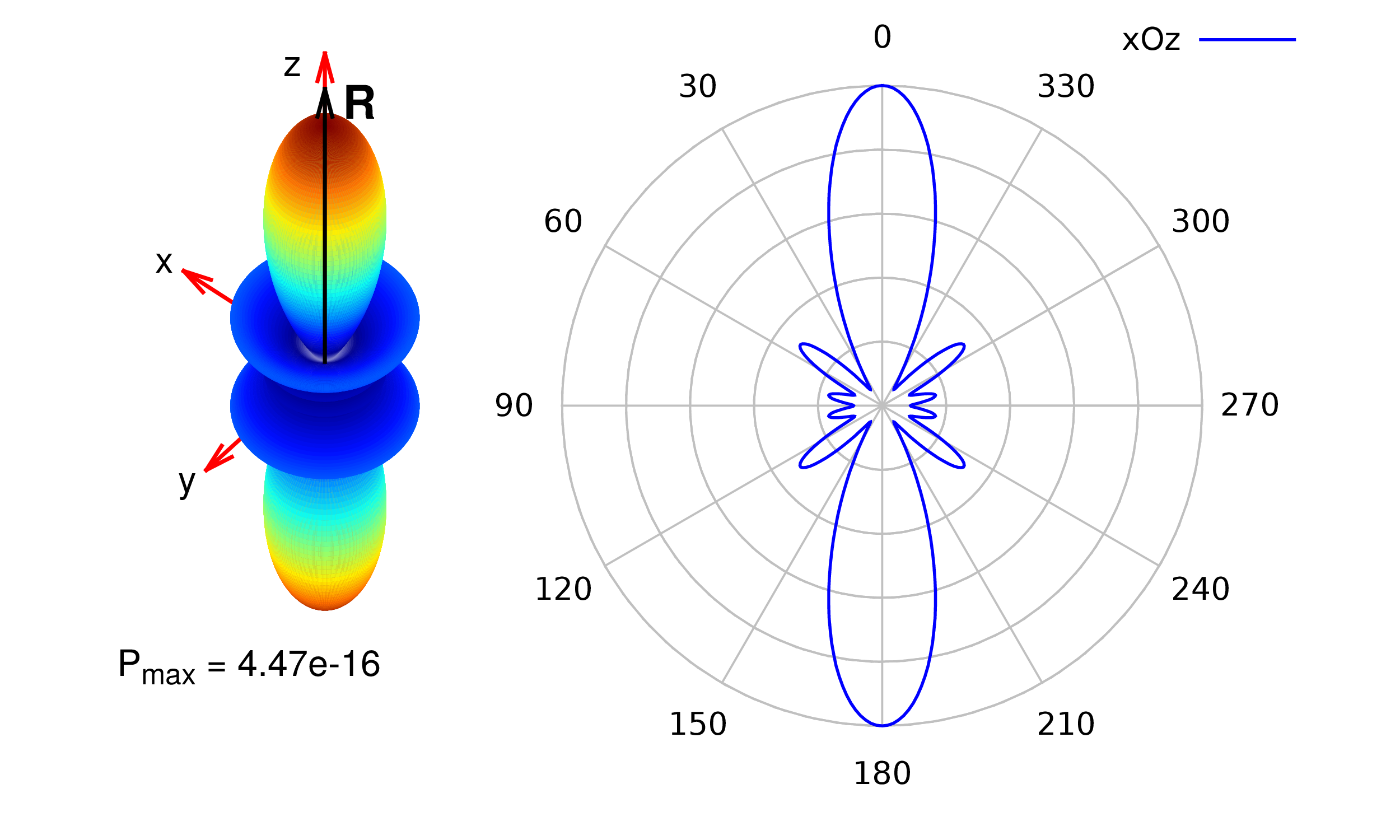}}
 \end{minipage}
 \caption{The ionization probability densities as a function of electron ejection
directions for fixed electron energies (E) chosen from
the multiphoton ionization peaks. The molecular axis (R) is fixed
along the polarization axis of the laser field (Oz axis), i.e. $\theta_R = 0^\circ$. The spherical
plots (left graph in each subfigure) are shown in the reference frame
defined in Figure \ref{fig:geometry}. On the polar plots (cuts along
the xOz plane), the ejection angle is measured from the Oz axis. The
maximum of each angular distribution (indicated with $P_{max}$ on
each subfigure) is normalized to 1 for easier comparison. }
\label{fig:sph_polar_00}
\end{figure*}

In order to better understand the dynamics behind the formation of the multiphoton peaks,
Figure (\ref{fig:sph_polar_00}) shows the angular distribution of the photoelectrons at fixed electron energies sampling 
the multiphoton peaks from Figure (\ref{fig:endep_00}). By inspection of the angular distributions, the dominant 
angular momentum of the ejected electrons in each multiphoton peak can be identified as $l=4$, $l=5$ and $l=6$ for the 
first, second and third multiphoton peak, respectively. Also, knowing that for $R < 3 $ a.u. internuclear separations, the 
contribution of the $l=0$ partial wave to the ground state electronic wave function is greater than $90\%$ leads to the 
conclusion that the first multiphoton peak is obtained after the absorption of a minimum number of 4 photons. 
Likewise, it can also be concluded that the second peak is a 5-photon process, while the third is a 6-photon one. 
It can also be observed that there is electron emission perpendicular to 
the molecular axis (in the $90^\circ$ and $270^\circ$ directions) for the first and the third multiphoton peaks indicating that there are continuum electrons only with 
even angular momentum. An even final state can only be reached from an even  ground state by the absorption of an even number of 
photons. Similarly, the lack of electron emission perpendicular to the molecular axis in the case of the second and fourth 
multiphoton peaks indicates a final state composed of odd partial waves and the absorption of an odd number of photons. 
These observations confirm that the order of each multiphoton peak has been correctly identified.

The fact that the 4-photon peak is located on top of the 3-photon non-sequential threshold may be initially misleading 
but it only indicates that the formation of the multiphoton peak is the result of a sequential process. The most probable 
scenario is where the molecule is initially excited from the $V_G$ to the $V_X$ potential energy surface via a single 
photon electronic transition. Then the molecule starts to dissociate on the $V_X$ potential energy surface, while electronic 
energy is being continuously transferred to the cores. Along this dissociation pathway, the electron may absorb further photons, 
promoting the system to the $V_I$ ionization potential energy surface and leading to the observed multiphoton spectrum. 
The shift of each multiphoton peak compared to its corresponding nonsequential threshold can be attributed to the 
transfer of electronic energy toward the nuclei. 

\begin{figure}
 \includegraphics[width=0.45\textwidth]{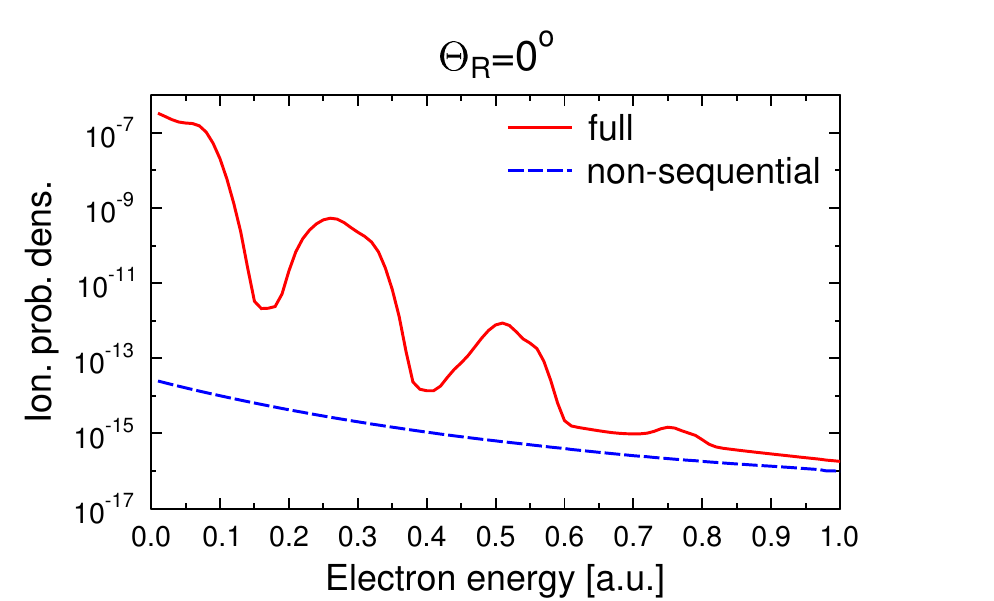}
 \caption{Same as Figure (\ref{fig:endep_00}). The results of the full calculations are compared to the results where only the non-sequential ionization pathways are allowed.}
\label{fig:endep_the0_nox}
\end{figure}

In order to test the above presented hypothesis, the nonsequential ionization spectrum has been calculated
by removing the $V_X$ potential energy surface from our model, thus
removing the sequential ionization pathway. Figure (\ref{fig:endep_the0_nox}) shows the results of these calculations 
where the electron ejection angle integrated ionization probability is shown as a function of electron energy. 
These results are compared with those of the full calculation where both the sequential and non-sequential ionization pathways are included.
On the figure it can be clearly observed that the contribution of the 
non-sequential ionization pathway (i.e. direct ionization from the $V_G$ potential energy surface) to the ionization spectrum is negligible 
and the dominant ionization pathway is the sequential one. 

\begin{figure}
 \includegraphics[width=0.45\textwidth]{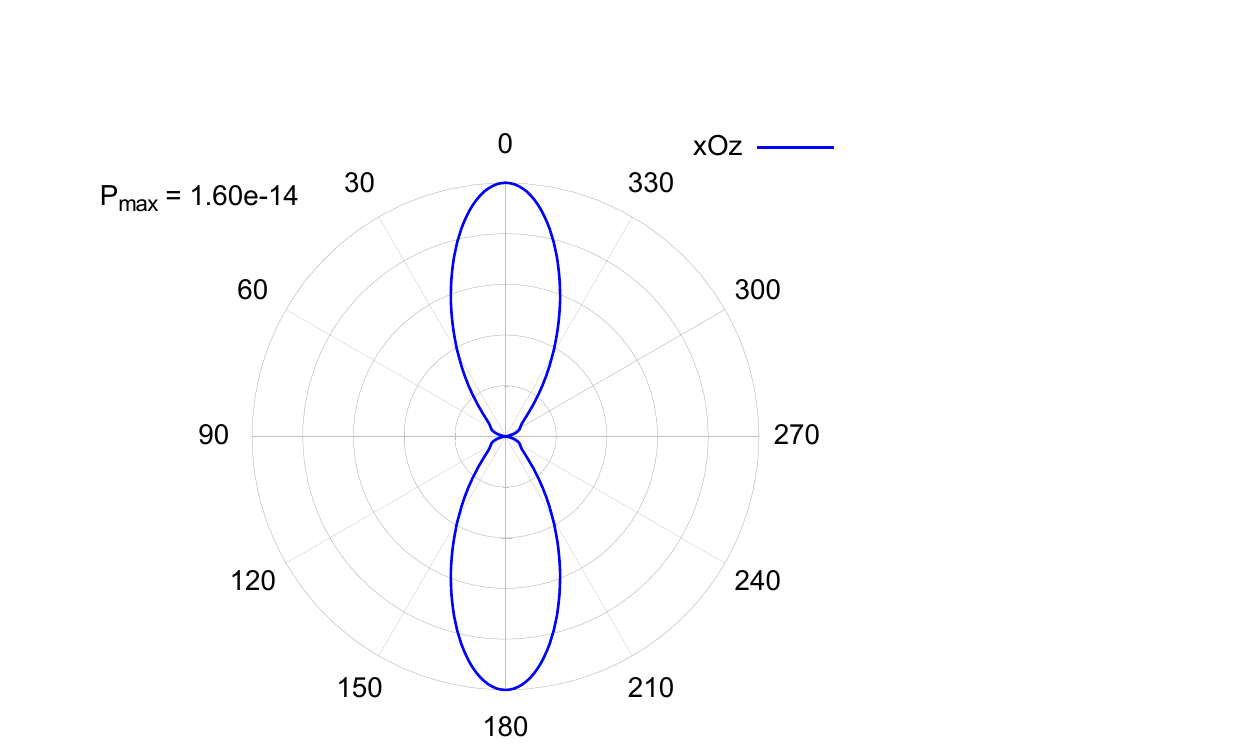}
 \caption{The angular distribution of the photoelectrons in the xOz plane at the $E=0.08$~a.u. electron energy. 
 The maximum of the angular distribution is normalized to 1, and the electron ejection angles are measured from the 
 laser polarization direction (Oz axis), which coincides with the molecular axis ($\theta_R = 0^\circ$).}
\label{fig:polar_the0_nox}
\end{figure}

It can also be observed on Figure (\ref{fig:endep_the0_nox}) that the shape of the full and the non-sequential 
electron spectrum are significantly different as there are no multiphoton peaks in the non-sequential curve. 
The source of this discrepancy can be understood from Figure (\ref{fig:polar_the0_nox}), which shows
the angular distribution of the photoelectrons at a fixed electron energy ($E=0.08$~a.u.). The observable dipole 
distribution indicates that for these laser field parameters the dominant angular momentum in the 
non-sequential pathway is $l=1$. This shows that the non-sequential ionization is the result of an one-photon 
transition and the multi-photon transitions contribution is masked by the dominant single-photon contribution. 
This happens due to the finite duration of the laser pulse used in our calculations, which has a wide spectral 
bandwidth with slowly decreasing tails.     

 \begin{figure}
 \includegraphics[width=0.45\textwidth]{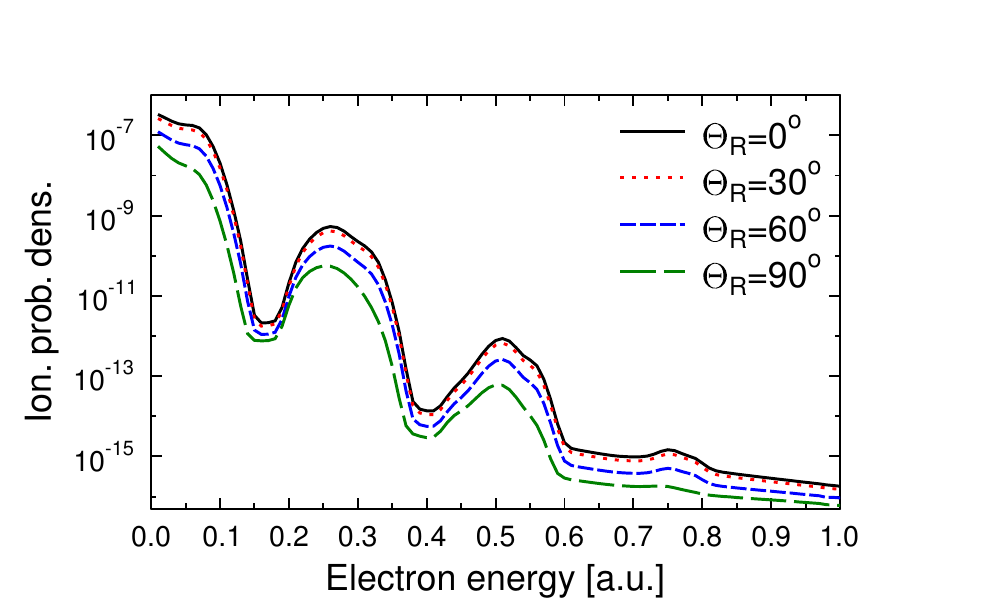}
 \caption{Angle integrated ionization probability density as a function of electron ejection
energy for different $\theta_{R}$ molecular axis orientations. The laser pulse parameters are $I_{0}$=3$\times$10$^{13}$
W/cm$^{2}$, $\tau$ = 10 fs.
 \label{fig:endep_thetardep}}
 \end{figure}

 \begin{figure}
 \includegraphics[width=0.45\textwidth]{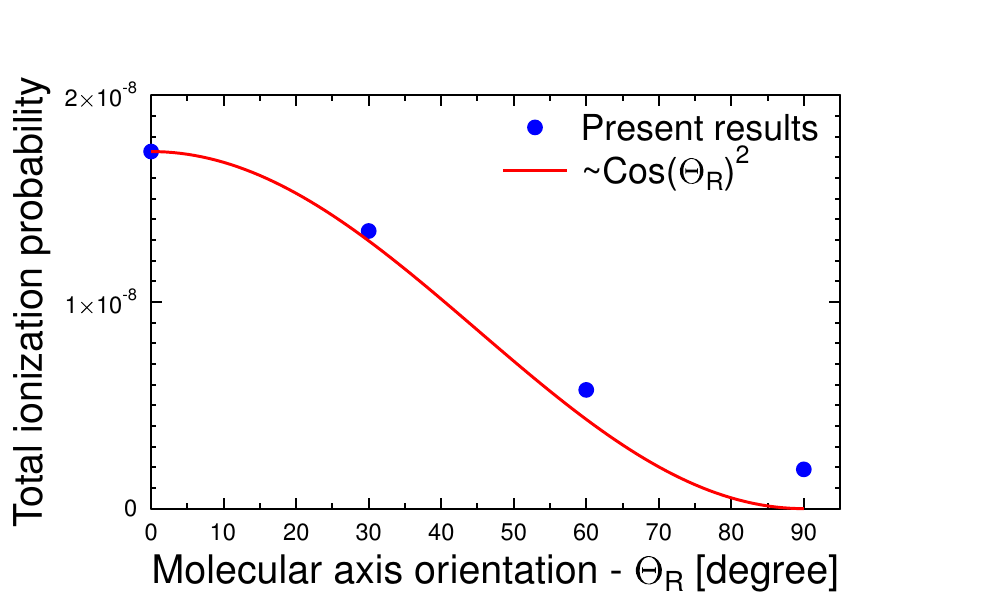}
 \caption{The total ionization probability as a function of the  $\theta_{R}$ molecular axis orientations (full circles) 
 are presented along with a normalized $\cos(\theta_R)^2$ curve. The laser pulse parameters are $I_{0}$=3$\times$10$^{13}$
W/cm$^{2}$, $\tau$ = 10 fs.
 \label{fig:angdep}}
 \end{figure}

Attention now turns toward the investigation of the molecular axis orientation dependence of the electron ejection spectra. 
Figure \ref{fig:endep_thetardep} presents the electron ejection angle integrated ionization probability density as a function of electron energy 
calculated for different $\theta_R$ molecular axis orientations. 
It can be observed that the multiphoton peaks are present in the photoelectron spectrum regardless of the molecular axis orientation.
In contrast to this, the total ionization probability density is strongly influenced by the $\theta_R$ molecular axis orientation. This dependence 
is shown in Figure \ref{fig:angdep} where the total ionization probability is presented as a function of the molecular axis orientation ($\theta_R$). 
The data can be interpreted by considering the transition dipole between a bound and a continuum electronic state which can be expressed as
\[
 \vec d(\vec k)=\vec d_\parallel(\vec k) + \vec d_\perp(\vec k),
\]
where $\vec d_\parallel(\vec k)$ is the component parallel to the molecular axis, while the $\vec d_\perp(\vec k)$ component is perpendicular to the 
molecular axis and is coplanar with the asymptotic electron momentum $\vec k$. For small molecular axis orientation angles, the laser coupling between 
the bound and continuum states is proportional to $\hat\varepsilon\vec d \simeq \hat\varepsilon\vec d_\parallel = d_\parallel\cos(\theta_R)$ and thus, 
the continuum states' population should also be proportional to $\cos(\theta_R)^2$. Figure \ref{fig:angdep} clearly shows that the 
total ionization probability closely follows the $\cos(\theta_R)^2$ curve for a large $\theta_R$ interval. This can be explained by the fact that 
$\vec d_\parallel \gg \vec d_\perp$, thus in the $\vec \varepsilon \vec d$ laser coupling, the $\vec \varepsilon \vec d_\perp$ term has significant 
contribution only when $\cos(\theta_R)$ is negligible, i.e. when the molecular axis is nearly perpendicular to the $\vec \varepsilon$ laser polarization axis.

\begin{figure*}
 \begin{minipage}{0.45\textwidth}
 \subfloat[$\theta_R=0^\circ$]{\includegraphics[width=\textwidth]{thr00_spherical_A.png}}
 \end{minipage}
 \begin{minipage}{0.45\textwidth}
 \subfloat[$\theta_R=30^\circ$]{\includegraphics[width=\textwidth]{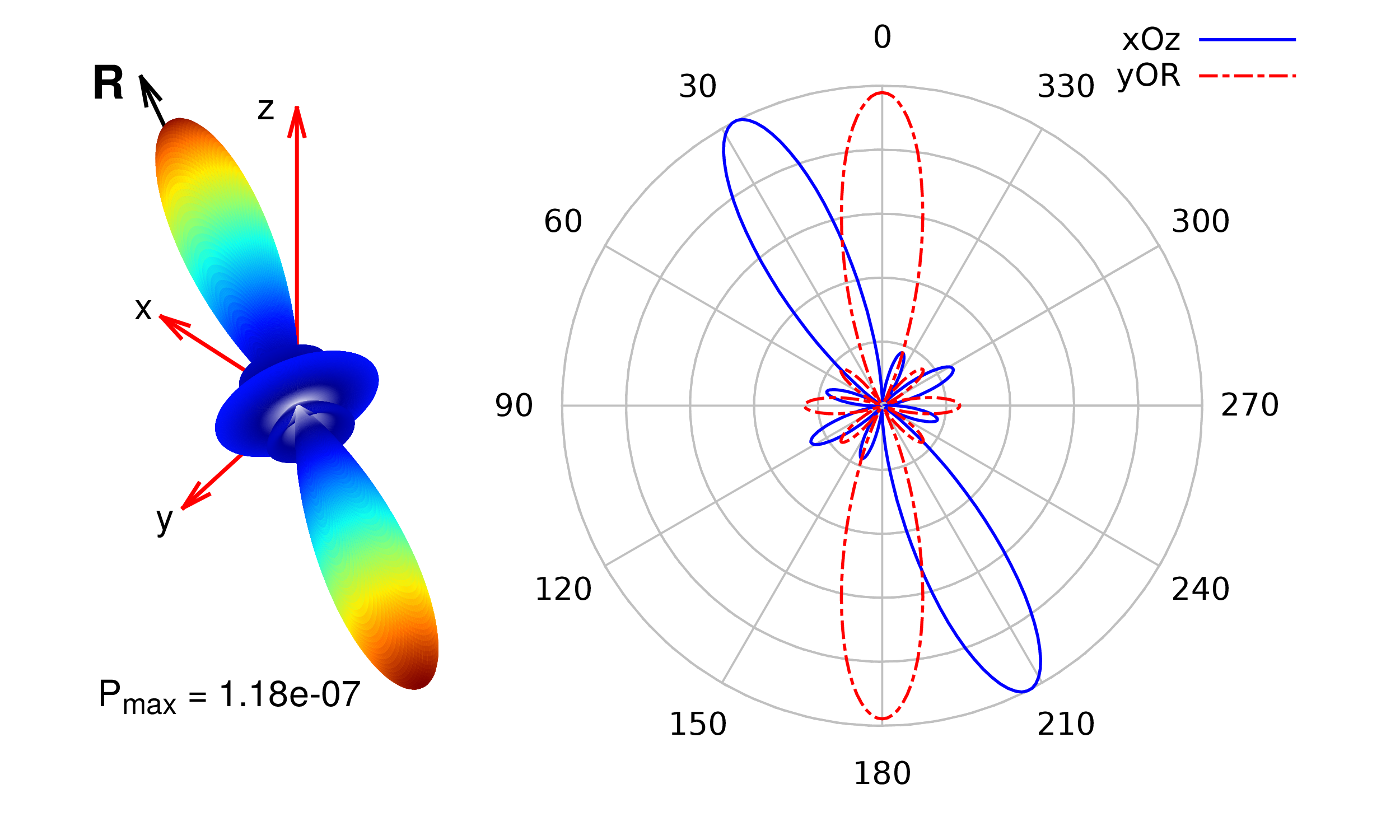}}
 \end{minipage}
  \begin{minipage}{0.45\textwidth}
 \subfloat[$\theta_R=60^\circ$]{\includegraphics[width=\textwidth]{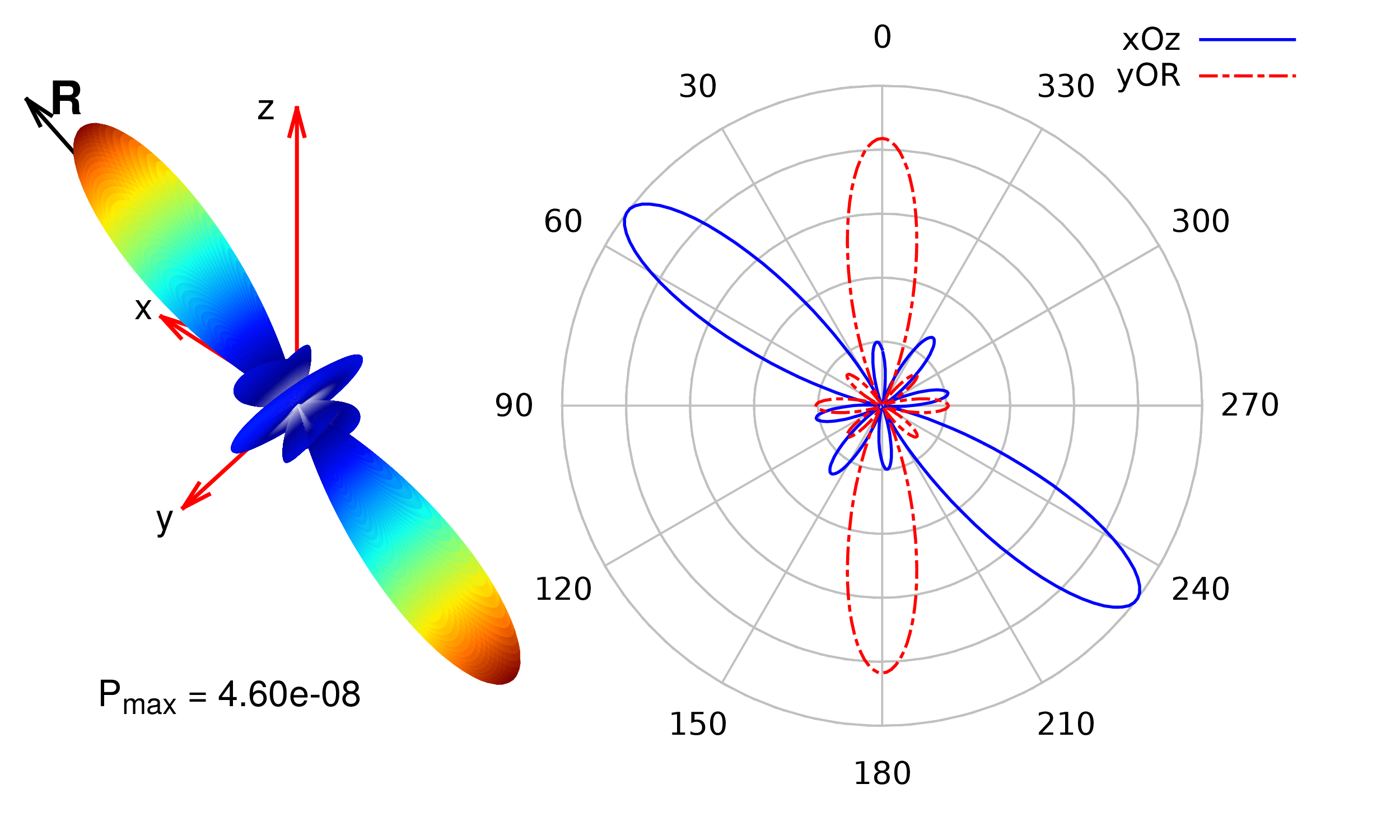}}
 \end{minipage}
  \begin{minipage}{0.45\textwidth}
 \subfloat[$\theta_R=90^\circ$]{\includegraphics[width=\textwidth]{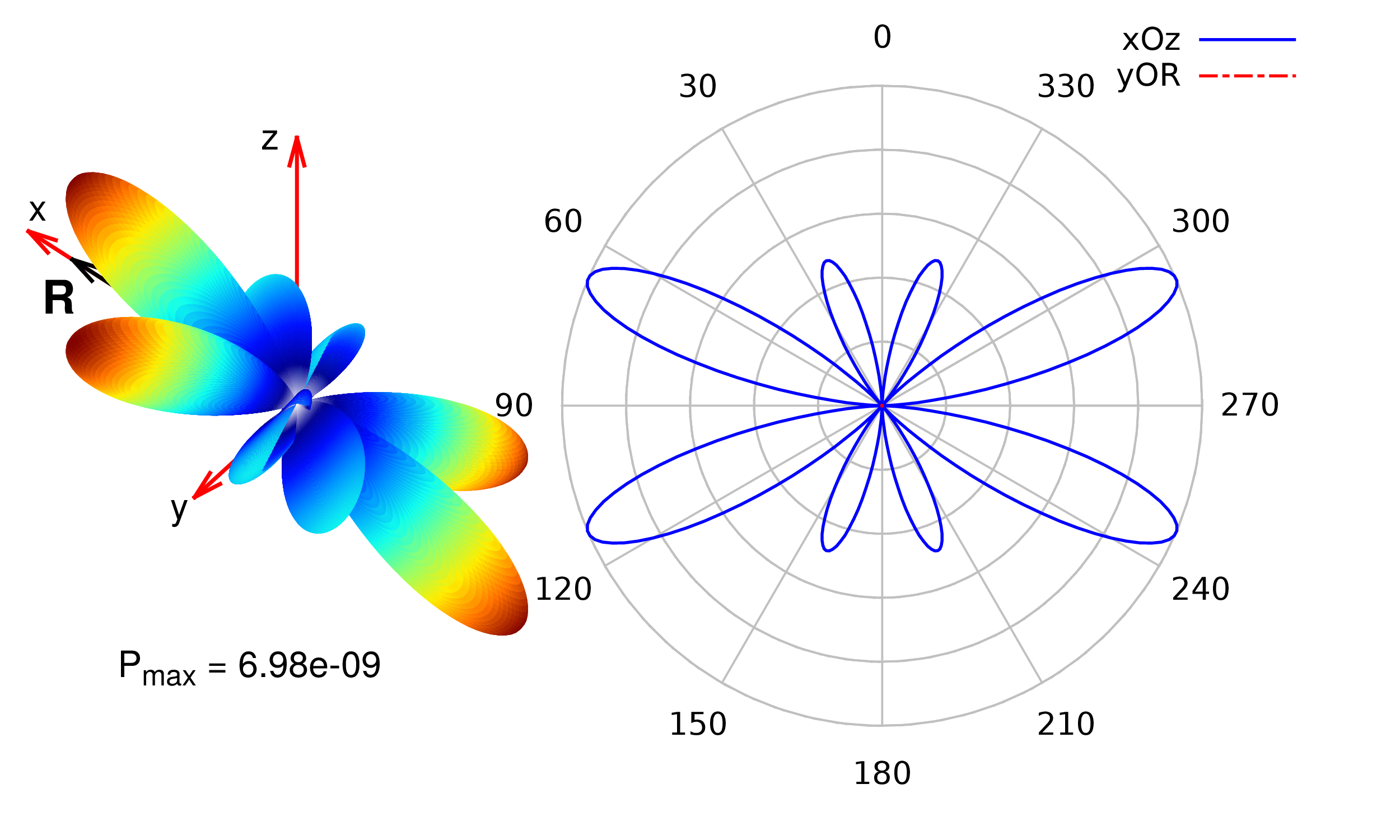}}
 \end{minipage} 
  \caption{The ionization probability densities as a function of electron ejection
directions are presented for a fixed E= 0.08 a.u. electron energy, and for different molecular axis orientations. The spherical
plots (left graph in each subfigure) are shown in the reference frame
defined in Figure \ref{fig:geometry}. On the polar plots cuts along
the xOz plane and yOR planes are presented, where the ejection angle is measured from the Oz axis (xOz) or from $\vec R$ the molecular axis (yOR). The
maximum of each angular distribution (indicated with $P_{max}$ on
each subfigure) is normalized to 1 for easier comparison.}
\label{fig:sph_Rdep}
\end{figure*}

The dominance of the $\hat\varepsilon \vec d_\parallel$ term in the laser coupling can also be observed in Figure \ref{fig:sph_Rdep} 
where the electron ejection angle dependent ionization probability at fixed electron energy (E=0.08 a.u.) is presented for different 
molecular axis orientations. For a large interval of molecular axis orientations ($0^\circ<\theta_R<60^\circ$), the shape of the angular 
distributions is defined by the $\hat\varepsilon\vec d_\parallel$ component of the laser coupling. In this interval, the obtained 
angular distributions for $\theta_R\ne 0^\circ$ molecular axis orientations are very similar to the one obtained for $\theta_R=0^\circ$. 
The angular distribution of the photoelectrons follows the rotation of the molecular axis without any major change in it's shape. 

Increasing $\theta_R$ results in the mirror symmetry of the angular distribution relative to the molecular axis in the xOz plane being lost, 
and the angular distribution tilts in the direction of the laser polarization axis. This effect is caused by $\hat\varepsilon \vec d_\perp$ 
which in the direction of the laser polarization axis increases whilst in the mirrored direction decreased the laser coupling.    
 
The limiting case of $\theta_R=90^\circ$ yields a completely different angular distribution than 
the other molecular axis orientations. In this case, the laser coupling is solely composed of the $\hat\varepsilon\vec d_\perp$ term. In contrast 
to the $\hat\varepsilon\vec d_\parallel$ term which couples the ground and excites states to continuum states with $m=0$ symmetry, the 
$\hat\varepsilon\vec d_\perp$ couples the ground and excited states to final states with $m\ne 0$. This leads to a continuum composed 
of partial waves with high $m$ magnetic quantum number and to the completely different angular distribution observed in Figure \ref{fig:sph_Rdep}(d). 
For the $\theta_R=90^\circ$ molecular axis orientation the electron emission in the yOR plane is forbidden since 
both the $\hat\varepsilon\vec d_\perp$ and $\hat\varepsilon\vec d_\parallel$ terms vanish in this plane.

Figure \ref{fig:sph_Rdep}  shows the molecular axis orientation dependence of the angular distribution of photoelectrons only 
for the first multiphoton peak. Similar behavior can be observed, and the same conclusions can be drawn for the other multiphoton peaks 
hence we omit their presentation here.
 
\subsection{The \ce{HeH++} system}

In order to benchmark our model, calculations have been performed for the ionization of the \ce{HeH++} by XUV laser pulses. 
and compared the results with the available  high precision \textit{ab initio} calculations \cite{Guan}, 
and with first order perturbation theory calcutations. Due to the high photon energy and moderate intensity of the used 
XUV pulse, the dominant ionization mechanism was single-photon ionization which can be described with high accuracy using 
first order perturbation theory.

\subsubsection{Technical details}
Our model was modified to replicate the available \textit{ab initio} data obtained for fixed molecular axis length and orientation.
This fixed-nuclei arrangement can be achieved in the simplest way by adding a
masking function to the $\vec d_{ij}(R,\theta_R)$ transition dipole moments:
\begin{equation}
S(R)=\begin{cases}
\begin{array}{cl}
1 & \mathrm{if}\  R \in (R_0-\Delta R,R_0+\Delta R)\\
0 & \mathrm{elsewhere} 
\end{array} \end{cases},
\label{eq:mask}
\end{equation}
where $R_0$ is the fixed molecular axis length. The effect of the $S(R)$ masking function is that it restricts the transition between 
the potential energy surfaces to a narrow vicinity of $R_0$.

These calculations were performed in a simulation box with $R_{max}=R_0+4.5$ a.u. size and 25 $\mathrm{a.u.}^{-1}$ gridpoint density. 
The CAP applied to the ground electronic state was placed at $R_{max} - 2$ a.u., while on the excited and ionization states a CAP with $\eta=0.25$ was placed at $R_{max} - 1.5$ a.u.
For each simulation run, the system was initialized at the ground potential energy surface as a narrow 
Gaussian wave packet centered around $R_0$ with a width of 0.05 a.u. The width of the Gaussian wave packets was chosen to prevent it's spreading
during the action of the laser pulse and it also perfectly fits in the $\Delta R = 0.1$  a.u. masking window.  

The electronic states and the transition dipole moments between them
were calculated by solving the electronic Schr\"{o}dinger equation directly
as outlined in subsection \ref{sub:The-electronic-Schr=0000F6dinger}.
The numerical convergence of the calculated $\Psi_{G}(R)$ and $\Psi_{X}(R)$
was carefully verified and the derived potential energy surfaces
were in excellent agreement with the \textit{ab initio} data \cite{Guan}.

First order time-dependent perturbation theory \cite{Bransden} calculations were also performed.
In this famework the fixed-nuclei transition probability from the ground state to an ionic state with a $\vec{k}$ continuum electronic momentum can be calculated as
\begin{equation}
 P^{1st}(\vec k, \vec R_0,t)=\left|\left\langle \psi_c(\vec k,\vec r)|\vec\varepsilon\vec r| \psi_G(\vec R_0,\vec r) \right\rangle\int\limits_0^tdt^\prime F(t^\prime)e^{[\frac{k^2}{2}-E_{G}(R_0)]t^\prime} \right|^2,
 \label{eq:pert}
\end{equation}
where all components (ground and continuum electronic wave functions, shape of the electric field) are already known. 
\subsubsection{Energy integrated angular distribution of photoelectrons} 
\begin{figure*}
 \includegraphics[width=0.95\textwidth]{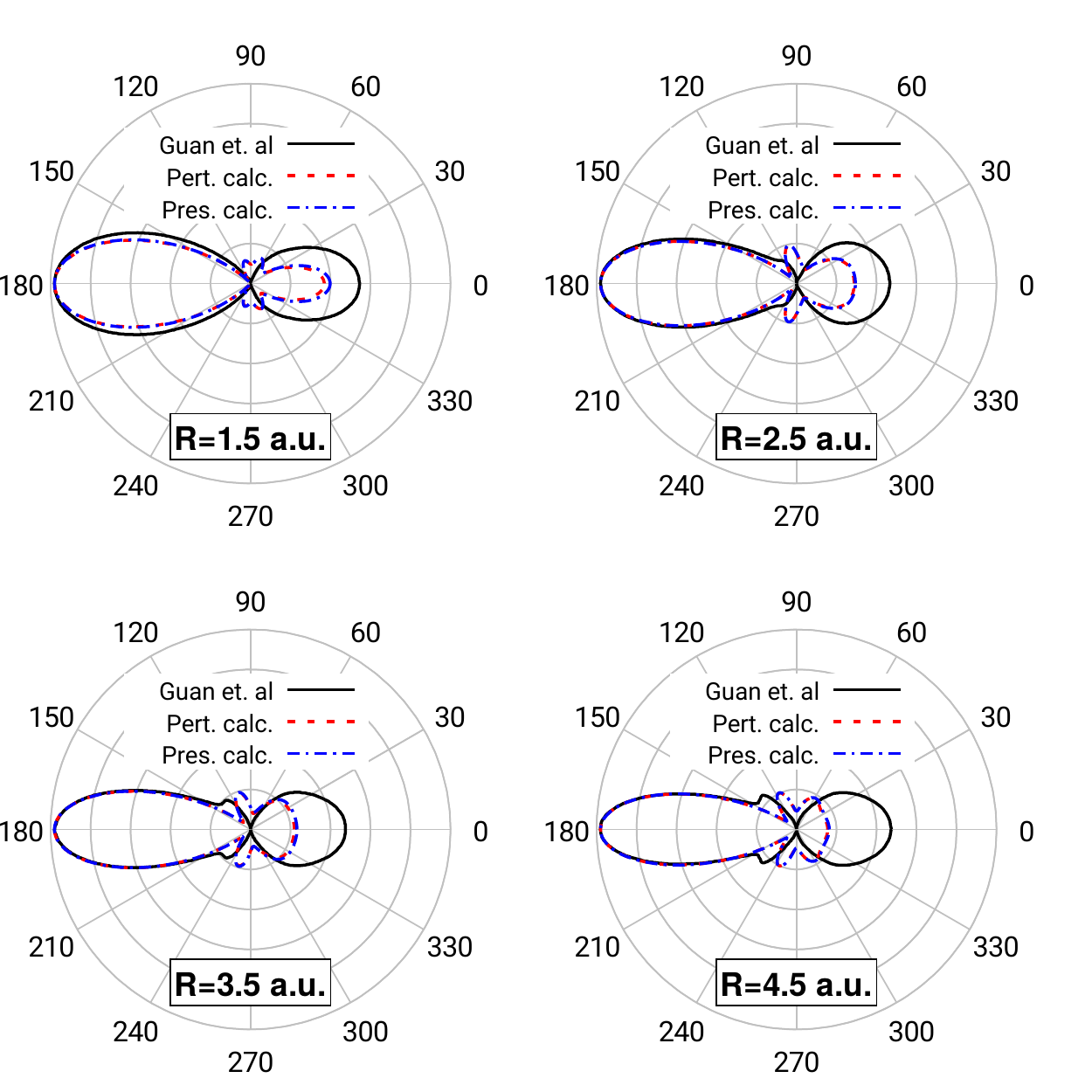}
 \caption{The electron momentum integrated angular distribution of photoelectrons emitted by \ce{HeH++} for different fixed molecular axis lengths. 
 The molecular axis orientation is also fixed in the direction of laser polarization vector. The results of the \textit{ab initio} and perturbative calculations
 are compared with the results of Guan et. al \cite{Guan}. The length of the laser pulse was fixed at 25 optical cycles, 
 the photon energy was 200 eV, while the intensity was set to $I_0=10^{13} \mathrm{W/cm}^2$.  
 \label{fig:HHE_angdep}}
 \end{figure*}
We have calculated the ionization probability density of the \ce{HeH++} molecular ion interacting with an XUV ultrashort laser pulse. 
In accordance to \cite{Guan} the photon energy was chosen to be 200 eV, the duration of the 
pulse was fixed at 25 optical cycles, while the intensity at $I_0=10^{13}$ W/cm$^2$. 
Figure \ref{fig:HHE_angdep} shows a good agreement of our results for different molecular axis lengths ($R$) with the reference data of Guan et. al \cite{Guan}
at the level of electron momentum integrated angular distribution of photoelectrons.
For an easier comparison all presented data are normalized to their maximum. 
In contrast to the \ce{D2+} situation where the behaviour of the photoelectron angular distributions as a function of molecular axis orientations
has been extensively investigated, this study is constrained to the case of molecular axis orientation parallel to the laser polarization axis.  
 Regardless of the molecular axis length the electron emission in all models is predominantly occurring in the direction 
 of the \ce{H} core ($\sim$ 180 degrees). The agreement between our present model and the first order perturbation calculations 
 with fixed nuclei is excellent. This is not surprising since both use the same initial and final electronic states. This 
 agreement suggests, that the inclusion of the masking function $S(R)$ (\ref{eq:mask}) in our model is a suitable approach 
 to reproduce fixed-nuclei calculations. There are also discrepancies between the reference \cite{Guan} and the present 
 \textit{ab initio} results. Our model underestimates the electron emission in the direction of the \ce{He} core and presents 
 a different shape of the angular distribution perpendicular to the molecular axis (between 60 and 120 degrees).  
 These discrepancies are assumed to be caused by the different description of the continuum electrons. In our model, simple 
 one-center Coulomb wave functions (see Eq. \ref{eq:coulomb}) centered at the charge center of the molecule are used while 
 Guan et. al \cite{Guan} uses the exact two-center Coulomb wave functions. This indicates that our model would be improved 
 by switching to the two-center Coulomb wave functions for the description of the continuum electrons.  
\section{Conclusions and Outlook}

In the present paper we have developed and tested a simple model for the \textit{ab initio} description of the combined electronic and nuclear 
motion of diatomics with a single active electron. In the framework of this model, we have followed the nuclear motion on electronic potential 
energy surfaces which have an important role in the dynamics. We have extended an existing approach 
\cite{Gabor1,Gabor2,Gabor3,Gabor4,Gabor5,Gabor6,Gabor7,Gabor8} which considered only bound electronic states in the dynamics by including 
the potential energy surfaces corresponding to ionized states. This way, beside the dissociation we were also able to study the ionization. 
Continuum-continuum transitions and the ground state depletion effects are neglected by the present model as the calculations for different 
ionization states are performed independently.

Within the framework of this model, we have performed test calculations for the multiphoton ionization of the \ce{D2+} molecule, 
and we have analyzed in details the resulting ionization spectra. The energy and angular dependence of the obtained data could be
explained using simple arguments.

The model was then benchmarked using the ionization of \ce{HeH++} in the single-photon regime and our results were compared to the reference data 
of Guan et. al \cite{Guan}. There was a good agreement between the two data sets, however small discrepancies were also identified pointing 
out possible ways of improving our approach. 

The reliability of our model in the single and multiphoton ionization regimes shown throughout these test calculations gives us a solid base 
for the further improvement of the model. In the next step, the kinetic energy release (KER) spectrum for both the dissociation and ionization 
channels will be calculated for fixed molecular axis orientations. The model will then be further improved by including the 
rotational degree of freedom as well. At that stage of development, we will be able to study how the LICI is influencing both the electronic and KER spectrum. 

\section*{Acknowledgment}

The authors acknowledge the financial support by a grant of the Romanian
National Authority for Scientic Research project number
PN-II-ID-PCE-2011-3-0192, by the European COST Action CM1204 XLIC and by
the TAMOP-4.2.2.B-15/1/KONV-2015-0001.  Á.V. and S. B. also acknowledge the OTKA
(NN103251 and K103917) projects.  Financial support by the Hungarian Academy of
Sciences is gratefully acknowledged. The authors thank Lorenz Cederbaum
for many valuable discussions. 

\bibliography{atoth_d2p_1_submit_1}

\end{document}